\documentclass[aps,prd,groupedaddress,twocolumn,showpacs,nofootinbib]{revtex4-1}
\usepackage{graphicx,epsfig}

\begin{document}


\title{Gamma-rays from ultracompact minihalos: potential constraints on the primordial curvature perturbation}



\author{Amandeep S. Josan}
\email[]{ppxaj1@nottingham.ac.uk}


\affiliation{School of Physics and Astronomy, University of Nottingham, University Park, Nottingham, NG7 2RD, UK}

\author{Anne M. Green}
\email[]{anne.green@nottingham.ac.uk}


\affiliation{School of Physics and Astronomy, University of Nottingham, University Park, Nottingham, NG7 2RD, UK}


\date{\today}

\begin{abstract}

  Ultracompact minihalos (UCMHs) are dense dark matter structures
  which can form from large density perturbations shortly after
  matter-radiation equality. If dark matter is in the form of Weakly
  Interacting Massive Particles (WIMPs),
  then UCMHs may be detected via their gamma-ray emission.  We
  investigate how the {\em{Fermi}} satellite could constrain the
  abundance of UCMHs and place limits on the power spectrum of the
  primordial curvature perturbation. Detection by {\em Fermi} would
  put a lower limit on the UCMH halo fraction. The smallest detectable 
  halo fraction, $f_{\rm UCMH} \gtrsim 10^{-7}$, is for 
  $M_{\rm UCMH} \sim 10^{3} M_{\odot}$.
  If gamma-ray emission from UCMHs is not detected, an
  upper limit can be placed on the halo fraction. The bound is tightest,
  $f_{\rm UCMH} \lesssim 10^{-5}$, for $M_{\rm UCMH} \sim 10^{5}
  M_{\odot}$. The resulting upper limit on the power
  spectrum of the primordial curvature perturbation in the event of 
  non-detection is in the range $\mathcal{P_R}
  \lesssim 10^{-6.5}- 10^{-6}$ on scales $k \sim 10^{1}-10^{6} \, {\rm
  Mpc}^{-1}$. This is substantially tighter than the existing
  constraints from primordial black hole formation on these scales,
  however it assumes that dark matter is in the form of WIMPs and
  UCMHs are not disrupted during the formation of the Milky Way halo.
\end{abstract}


\maketitle


\section{Introduction}

The power spectrum of the primordial curvature perturbation on large
scales has been accurately measured using cosmological
observations~\cite{Komatsu:2010fb}. These measurements can be used to
constrain models of inflation (see e.g. Ref.~\cite{Peiris:2006sj}).
Cosmological observations only probe a very narrow range of scales,
however avoiding the overproduction of Primordial Black Holes (PBHs)
constrains the primordial perturbations over a wide range of smaller
scales.  If the density perturbation at horizon entry in a given
region exceeds a threshold value, $\delta_{\rm c}\sim 0.3$, then
gravity overcomes pressure forces and the region collapses to form a
PBH with mass of order the horizon mass~\cite{Carr:1974nx}. The
abundance of PBHs formed is constrained by the consequences of their
evaporation and their gravitational effects (for recent compilations
and updates of the constraints see
Refs.~\cite{Josan:2009qn,Carr:2009jm}).  These abundance constraints
can be translated into constraints on the power spectrum of the
primordial curvature perturbation of order $\mathcal{P_R} <
10^{-1}-10^{-2}$~\cite{Josan:2009qn}.

Ricotti \& Gould~\cite{Ricotti:2009bs}
have recently proposed that slightly smaller perturbations, in the
range $10^{-3}$ to $ \delta_{\rm c}$, can collapse before $z
\sim 1000$ and seed the formation of
ultracompact minihalos (UCMHs). Due to their early formation, the
central regions of UCMHs would have a high dark matter (DM) density. If
DM is
in the form of Weakly Interacting Massive Particles (WIMPs), WIMP
annihilation within UCMHs may lead to an observable gamma-ray
signal~\cite{Ricotti:2009bs,Scott:2009tu}.

Scott \& Sivertsson~\cite{Scott:2009tu} have investigated gamma-ray emission
from UCMHs formed from perturbations which enter the horizon at three
different epochs in the early Universe: ${\rm{e}}^{+}{\rm{e}}^{-}$
annihilation, and the QCD and electroweak (EW) phase transitions.
They find that an UCMH corresponding to the ${\rm{e}}^{+}{\rm{e}}^{-}$
annihilation epoch, which has present day mass $M_{\rm UCMH}(z=0) \sim
10^{2} M_{\odot}$, could be detected by the {\em Fermi}
satellite or current Air Cherenkov telescopes (ACTs), at a distance of $100 \, {\rm pc}$. If $1\%$ of the DM is in the form of UCMHs with
this mass there would be $\sim 3$ UCMHs within $100 \, {\rm pc}$ of the Earth~\cite{Scott:2009tu}. UCMHs formed at earlier epochs would be lighter, and hence more challenging to detect.

We extend this work and examine the constraints which would be placed
on the primordial curvature perturbation power spectrum by the possible
detection of UCMHs.  It has been shown that there are single field models 
of inflation which are compatible with cosmological observations and where 
the perturbation amplitude on small-scales is large enough to produce a 
significant density of PBHs~\cite{Peiris:2008be,Josan:2010cj} (see also 
references therein).  It is therefore possible that UCMHs may form from 
perturbations generated by single field slow roll inflation.
Phase transitions~\cite{Ricotti:2009bs,Scott:2009tu} or features in the inflationary potential ~\cite{Berezinsky:2010kq} could also lead to enhanced perturbations on small scales.
In this paper we do not fix the UCMH mass or abundance.  Instead we 
calculate the constraints on the UCMH halo fraction which would arise from the detection (or non-detection) of gamma-rays from UCMHs by {\em Fermi} as a function of UCMH mass. We then translate the UCMH
abundance constraints into constraints on the power spectrum of the
primordial curvature perturbation, as a function of scale.  In
Sec.~\ref{sectionformemit} we summarize the calculation of the
properties of the UCMHs and the resulting gamma-ray flux, following
Scott and Sivertsson. We then calculate the lower bound on the UMCH
halo fraction which would result from detection of an UCMH by {\em Fermi}. 
We also calculate the upper bound which would result if no UCMHs are detected.
In Sec.~\ref{sectionpowerspectrum} we outline the calculation of the
amplitude of the density contrast (c.f. Ref.~\cite{Josan:2009qn}),
and translate the potential constraints on the abundance of UCMHs into
constraints on the power spectrum of the primordial curvature
perturbation. We conclude with discussion in
Sec.~\ref{sectionconclude}.

\section{UCMH formation and gamma-ray emission}
\label{sectionformemit}

Ricotti \& Gould~\cite{Ricotti:2009bs} find that a density
  perturbation with amplitude at horizon crossing $\delta >10^{-3}$
  will grow sufficiently during radiation domination that it collapses
  at $z \geq 1000$, seeding the formation a UCMH which then grows via
  spherical infall.
It has been argued that PBHs can also seed the formation of
minihalos~\cite{Mack:2006gz,Ricotti:2007au,Ricotti:2009bs}, and 
the resulting gamma-ray emission
(assuming that the remainder of the dark matter is in the form of WIMPs) leads to
constraints on the abundance of
PBHs~\cite{Lacki:2010zf}, however we do not pursue
that possibility here.

At matter-radiation equality the DM mass within a UCMH forming region, $M(z_{\rm eq})$, is given by~\cite{Scott:2009tu}
\begin{equation}
M(z_{\rm eq})=f_{\chi} \left(\frac{1+z_{\rm eq}}{1+z_{\rm i}} \right) M_{\rm H}(z_{\rm i})  \,,
\end{equation}
where
$f_{\chi}=\Omega_{\rm{DM}}/\Omega_{\rm{m}}=0.834$~\cite{Komatsu:2010fb}
is the dark matter fraction and $M_{\rm{H}}(z_{\rm i})=(4 \pi/3)\rho H^{-3} $ is the
horizon mass at redshift $z_{\rm i}$ corresponding to the epoch when the scale of interest entered the horizon.
After matter-radiation equality the UCMH mass, $M_{\rm
UCMH}(z)$, grows, due to radial infall of matter, as
\begin{equation}
M_{\rm{UCMH}}(z)=M(z_{\rm eq}) \left(\frac{1+z_{\rm{eq}}}{1+z} \right) \,.
\label{halomass}
\end{equation}
Following Scott \&
Sivertsson~\cite{Scott:2009tu} we assume that UCMHs stop growing at $z
\approx 10$ as the onset of structure formation prevents further
matter infall. Using the constancy of the
entropy, $s=g_{\star s}a^3T^3$, and the the radiation density,
$\rho=(\pi^2/30)g_{\star}T^4$, where $g_{\star s}$ is
the number of entropy degrees of freedom and $g_{\star}$ the number
of relativistic degrees of freedom and $T$ the temperature, the horizon mass can be
written as
\begin{equation}
M_{\rm{H}}(T)=M_{\rm{H}}(T_{\rm{eq}}) \left(\frac{g^{\rm{eq}}_{\star}}{g_{\star}}\right)^{1/2} \left(\frac{T_{\rm{eq}}}{T}\right)^2   \,.
\label{horizonmassT}
\end{equation}
Using $T \propto g_{\star s}^{-1/3} (1+z)$, the horizon mass as a function of redshift is given by
\begin{equation}
M_{\rm{H}}(z_{\rm i})=M_{\rm{H}}(z_{\rm{eq}}) \left(\frac{g^{\rm i}_{\star}}{g^{\rm eq}_{\star}}\right)^{1/6} \left(\frac{1+z_{\rm{eq}}}{1+z_{\rm i}}\right)^2   \,,
\label{horizonmassz}
\end{equation}
where we have taken $g_{\star s} \approx g_{\star}$.

The UCMH dark matter density profile is given,  in the radial
  infall model,
by~\cite{Ricotti:2009bs,Scott:2009tu}
\begin{equation}
\rho_{\rm UCMH}(r,z)=\frac{3f_{\chi}M_{\rm{UCMH}}(z)}{16\pi R_{\rm{UCMH}}^{\frac{3}{4}}(z) r^{\frac{9}{4}}}   \,,
\label{ucmhdmprofile}
\end{equation}
where $R_{\rm{UCMH}}(z)$ is the radius of the UCMH at redshift $z$, given by
\begin{equation}
\left( \frac{R_{\rm{UCMH}}(z)}{\rm{pc}} \right) = 0.019 \left( \frac{1000}{1+z} \right)
\left( \frac{M_{\rm{UCMH}}(z)}{M_{\odot}} \right)^\frac{1}{3}  \,,
\label{ucmhsize}
\end{equation}
where $M_{\odot}$ is the mass of the sun.

Baryonic infall may lead to adiabatic contraction of the UCMH density
profile~\cite{Blumenthal:1985qy}.  Scott \& Sivertsson considered a variable fraction of the total UCMH mass condensing to form a constant density baryonic core. The dark matter density in the centre of the halo does not rise significantly and hence the change in the resulting gamma-ray flux is relatively small~\footnote{This is true for dark matter in the form of standard WIMPs, with the canonical annihilation cross-section deduced from
the measured dark matter density. Motivated by recent electron data, Scott and Sivertsson also considered a model with enhanced annihilation cross-section. In that case WIMP annihilation leads to a larger constant density core and adiabatic contraction then has a larger effect.}.
Given the uncertainties in the calculation we therefore do not consider adiabatic
contraction.

WIMP annihilation reduces the density in the inner regions of the
UCMH.  We use the standard estimate of the maximum density,
$\rho_{\rm{max}}$,~\cite{Ullio:2002pj,Scott:2009tu}
\begin{equation}
 \rho_{\rm{max}} \approx \frac{m_{\chi}} {\left<\sigma v\right> (t_{0}-t_{\rm{i}})}     \,,
\label{rhomax}
\end{equation}
where $m_{\chi}$ is the WIMP mass, $\left<\sigma v\right>$ the thermally
averaged product of the WIMP annihilation cross-section and speed,
$t_{0} \approx 13.7 \, {\rm{Gyr}}$~\cite{Komatsu:2010fb} the current
age of the Universe and we take the UCMH formation time as
$t_{\rm{i}}(z=z_{\rm{eq}}) \approx 77\, {\rm{kyr}}$~\cite{Kolb:1990vq}. The
UCMH present day density profile is thus given by $ \rho_{\rm UCMH}(r)
= {\rm min} \left\{ \rho_{\rm max} \,, \rho_{\rm UCMH}(r, z=10)
\right\}$, where $\rho_{\rm UCMH}(r, z=10)$ is given by
eq.~(\ref{ucmhdmprofile}).

The gamma-ray flux above a threshold energy $E_{\rm th}$,
$\Phi_{\gamma}(E_{\rm th})$, from WIMP annihilation within an UCMH
at a distance $d$ from the Earth can be written as
\begin{equation} 
\Phi_{\gamma}(E_{\rm{th}})=\frac{\Phi_{\rm{astro}} \Phi_{\rm{particle}}}{2d^2}   \,.
\label{Phi}
\end{equation}
The particle physics term, $\Phi_{\rm particle}$, is given by
\begin{equation}
\Phi_{\rm{particle}}=\frac{1}{m^2_{\chi}} \sum_{f} \int_{E_{\rm{th}}}^{m_{\chi}}
 \left<\sigma_f v \right> 
 \frac{{\rm d}N_{f}}{{\rm d}E} {\rm d} E  \,,
\label{Phiparticle}
\end{equation}
where $\sigma_{f}$ is the annihilation cross-section and ${\rm d} N_{f}/{\rm d} E$ the differential photon yield of the $f$th annihilation channel.
We use DarkSUSY~\cite{Gondolo:2004sc} to carry out a scan of the
  parameter space of the Minimal Supersymmetric Standard Model
and compute $\Phi_{\rm{particle}}$ for the sets of parameters
that are compatible with accelerator bounds and produce a present day
DM density compatible with the WMAP measurement of the DM density. When calculating the 
lower limit on the halo fraction of UCMHs which would arise from a
detection by {\em Fermi} we use the largest value of
$\Phi_{\rm{particle}}$ obtained from the DarkSUSY scan. Conversely when calculating the upper limit which would result
if no UCMHs are detected we use the smallest value.
The astrophysical factor, $\Phi_{\rm astro}$, is given by
\begin{equation}
\Phi_{\rm{astro}}=\int_{0}^{R_h} r^2 \rho_{\rm UCMH}^2(r, z=10) \,  {\rm d}r  \,.
\label{Phiastro}
\end{equation}

The {\em{Fermi}} point source sensitivity above 
$100\, \rm{MeV}$ is~\cite{NASA:website}
\begin{equation}
\Phi_{\gamma}(100\, \rm{MeV})=6\times10^{-9} \rm{cm}^{-2} \rm{s}^{-1} \,.
\label{pointsouce}
\end{equation}
For a given UCMH mass, $M_{\rm UCMH}(z=0)$, we determine the distance
$d$ within which a UCMH of this mass would be detectable at threshold sensitivity by
{\em{Fermi}}.  We then calculate the fraction of the Milky Way in the form of
UCMHs if there is a single UCMH within this distance. This is the
smallest UCMH halo fraction which could be detected by {\em Fermi}.
To do this we assume that the fraction of the DM in the form of UCMHs
is independent of position so that the local and global UCMH fractions 
are identical
\begin{eqnarray}
f_{\rm{UCMH}} \equiv \frac{\Omega_{\rm{UCMH}}}{\Omega_{\rm{DM}}}       \nonumber
&=& \frac{n_{\rm{UCMH,MW}}(r) M_{\rm{UCMH}}(z=0)} {\rho_{\rm{DM,MW}}(r)}     \\
&=& \frac{M_{\rm{UCMH}}(z=0)} {M_{\rm{DM,MW}}(<d)} \,,
\label{globallocal}
\end{eqnarray}
where $\rho_{\rm{DM,MW}}(r)$ is the density profile of the Milky Way halo, $n_{\rm{UCMH,MW}}(r)$ the number density of UCMHs  and $M_{\rm{DM,MW}}(<d)$ the mass of DM within a sphere of radius $d$ centred on the Earth. We assume a NFW~\cite{Navarro:1996gj} 
density profile for the Milky Way with parameters as found by Klypin et al.~\cite{Klypin:2001xu}.

\begin{figure}
\includegraphics[width=9cm]{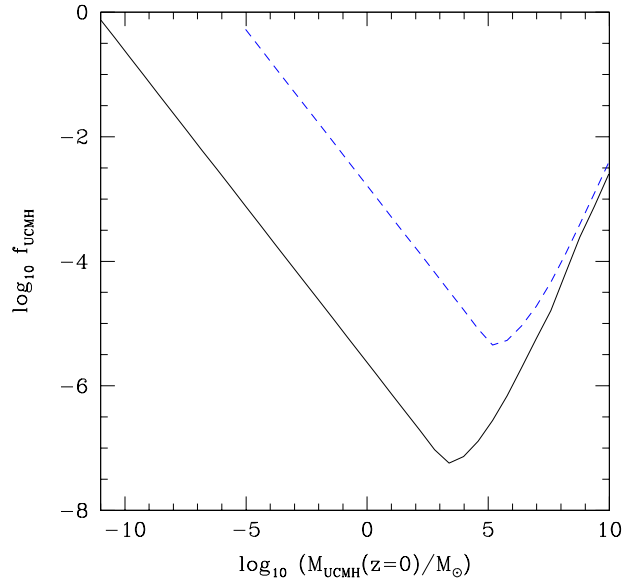}
\caption{Constraints on the UCMH halo fraction, $f_{\rm UCMH}$, 
  as a function of present day UCMH mass, $M_{\rm UCMH}(z=0)$.
  The solid line shows the lower bound on the halo 
  fraction which would result from the detection of gamma-rays from an UCMH by 
  {\em{Fermi}}. The dotted line shows the upper limit on
  the halo fraction if gamma-rays from UCMHs are not detected, assuming DM 
  is in the  form of WIMPs.}
\label{results1}
\end{figure}

Fig.~\ref{results1} shows the lower limit on the UCMH halo fraction, as a function of UCMH mass, which would result from the detection of a single UCMH by {\em Fermi} at threshold sensitivity. It also shows the upper limit on the UCMH halo fraction if {\em{Fermi}} does not detect gamma-rays from UCMHs, assuming that the DM is in the form of self-annihilating WIMPs. More massive UCMHs have 
a larger gamma-ray flux ($\Phi_{\rm astro} \propto M_{\rm UCMH}(z=0)$ roughly) and hence can be
detected at a larger distance ($d \propto M_{\rm UCMH}(z=0)^{1/2})$. For $M_{\rm UCMH}(z=0) \lesssim 10^{3} M_{\odot}$, $d \lesssim 10 \, {\rm kpc}$ so that $M_{\rm DM,MW}(<d)$ increases more rapidly than $M_{\rm UCMH}(z=0)$ 
resulting in a decreasing limit on the halo fraction as $M_{\rm UCMH}(z=0)$ is increased.
For more massive UCMHs $d$ becomes significantly larger than the scale radius of the Milky Way halo, hence $M_{\rm DM,MW}(<d) \propto \ln{[M_{\rm UCMH}(z=0)]}$ resulting in a subsequent increase in the limit on the halo fraction for $M_{\rm UCMH}(z=0) \gtrsim 10^{3} M_{\odot}$.

\section{Constraints on the primordial curvature perturbation}
\label{sectionpowerspectrum}

To translate the limits on the UCMH halo fraction into constraints on the primordial curvature perturbation, we need to
relate the present day UCMH halo fraction to the primordial density
perturbation distribution. The present day UCMH density,
$\Omega_{\rm{UCMH}}$, is related to the UCMH halo fraction, $f_{\rm
UCMH}$, by eq.~(\ref{globallocal}).  Assuming that UCMHs are not
destroyed by dynamical processes during structure formation, the
present UCMH density is related to the fraction of the Universe at 
horizon entry which is overdense enough to later form UCMHs, $\beta_{\rm{UCMH}}$,
by
\begin{equation}
\Omega_{\rm{UCMH}} = \Omega_{\rm DM} \frac{M_{\rm UCMH}(z=0)} {M(z_{\rm eq})} \,
\beta_{\rm{UCMH}}(M_{\rm H}(z_{\rm i})) \, 
\label{omegatobeta}
\end{equation}
As UCMHs are far more compact and dense than typical DM halos
they will be far less susceptible to disruption.  Our lower bounds are conservative; if UCMHs are destroyed, the initial abundance of UCMH forming perturbations, and hence the amplitude of the primordial perturbations, will be under-estimated. The upper limit from non-detection would, however, be weakened.

If the smoothed density contrast, in the comoving gauge, $\delta_{\rm hor}(R)$, at horizon crossing ($R= (a H)^{-1}$), is in the range $10^{-3} \leq\delta_{\rm hor}(R)\leq 1/3$, the DM in the region will eventually collapse to form an UCMH~\cite{Ricotti:2009bs}.  The horizon mass $M_{\rm{H}}(z_{\rm i})$ is related to the smoothing scale, $R$, by~\cite{Green:2004wb}

\begin{equation}
M_{\rm H}(z_{\rm i}) = M_{\rm H}(z_{\rm{eq}}) (k_{\rm{eq}} R)^2 \left(\frac{g^{\rm{eq}}_{\star}} {g^{\rm i}_{\star}} \right)^{\frac{1}{3}} 
\label{smoothingscale}
\end{equation}
where $k_{\rm{eq}}=0.07(\Omega_{\rm m} h^2) {\rm{Mpc}^{-1}}$ is the wavenumber and 
$M_{\rm H}(z_{\rm{eq}})= 1.3 \times 10^{49} (\Omega_{\rm m}
h^2)^{-2} \, {\rm g}$ the horizon mass at matter-radiation equality and we use $g^{\rm{eq}}_{\star} \approx 3$ and $g^{\rm i}_{\star} \approx 100$~\cite{Kolb:1990vq}.

The fraction of the Universe in regions dense enough to eventually form
UCMHs is given by Press-Schechter theory~\cite{Press:1973iz},
\begin{equation}
\beta_{\rm{UCMH}}(M_{\rm H}(z_{\rm i}))= 2
      \int_{10^{-3}}^{1/3} P(\delta_{\rm{hor}}(R))  {\rm d} \delta_{\rm{hor}}(R) \,,
\label{presssch}
\end{equation}
where, assuming that the initial perturbations are Gaussian, the probability
distribution of the smoothed density contrast, $P(\delta_{\rm
hor}(R))$, is given by (e.g. Ref.~\cite{Liddle:2000cg})
\begin{equation}
P(\delta_{\rm hor}(R))=\frac{1}{\sqrt{2\pi} \sigma_{\rm hor}(R)} \exp{ \left(
    - \frac{\delta_{\rm hor}^2(R)}{2 \sigma_{\rm hor}^2(R)} \right)} \,,
\end{equation}
where $\sigma(R)$ is the mass variance
\begin{equation}
\label{variance}
\sigma^2(R)=\int_{0}^{\infty} W^2(kR)\mathcal{P}_{\delta}(k, t)\frac{{\rm d}k}{k} \,,
\end{equation}
and $W(kR)= \exp{(-k^2 R^2/2)}$ is the Fourier transform of the window function used to
smooth the density contrast, which we take to be Gaussian.
The relationship between the present UCMH density and the mass variance is then given by
\begin{eqnarray}	
\Omega_{\rm UCMH}& \approx & \frac{2 \Omega_{\rm DM}}{\sqrt{2\pi}\sigma_{\rm hor}(R)} 
\frac{M_{\rm UCMH}(z=0)}{M(z_{\rm eq})} \,,
 \nonumber
   \\
&\times& \int_{10^{-3}}^{1/3} \exp{\left(- \frac{\delta^2_{\rm hor}(R)}
{2 \sigma_{\rm hor}^2(R)}\right)} {\rm d} \delta_{\rm hor}(R) \,.
\label{densitypara}
\end{eqnarray}
The constraints on the present day UCMH density can therefore be
translated into constraints on the mass variance by simply inverting
this expression.  The final ingredient required to calculate
constraints on the power spectrum of the primordial curvature perturbation is the relationship between the density contrast in the
comoving gauge and the primordial curvature perturbation, ${\cal R}$.  This has recently been calculated taking into account the evolution of perturbations prior to horizon entry (see Ref.~\cite{Josan:2009qn} for details):
\begin{equation}
\delta(k,t)=-\frac{4}{\sqrt{3}} \left(\frac{k}{aH}\right) 
     j_{1}(k/\sqrt{3}aH) \mathcal{R} \,,
\end{equation}
where $j_{1}$ is a spherical Bessel function.
The power spectrum of the density contrast is therefore given by:
\begin{equation}
\mathcal{P}_{\delta}(k,t)=\frac{16}{3} \left(\frac{k}{aH}\right)^2 
    j_1^2(k/\sqrt{3} aH)
       \mathcal{P}_{\mathcal{R}}(k) \,.
\label{powerallscales}
\end{equation}
Substituting this into eq.~(\ref{variance}), and setting $R=(a H)^{-1}$, gives
\begin{eqnarray}
\sigma_{\rm hor}^2(R) &=& \frac{16}{3} \int_{0}^{\infty}
     \left( kR \right)^2  j_{1}^2(k R/\sqrt{3}) \nonumber \\
  && \times
 \exp(- k^2 R^2)
         \mathcal{P}_{\mathcal{R}}(k)  \frac{{\rm d} k}{k}\,.
\label{variancefinal}
\end{eqnarray}

This integral is dominated by scales $k\sim k_{0} = 1/R$. Following
Ref.~\cite{Josan:2009qn} in the context of slow-roll inflation models
we can assume that the power spectrum is constant over these scales,
${\cal{P}}_{\cal{R}} (k) = {\cal{P}}_{\cal R}(k_0) $.  Relaxing this 
assumption and using a power law power spectrum with spectral index in 
the range consistent with slow-roll inflation, $0.9<n(k_0)<1.1$, leads to 
changes of order $3\%$ in the power spectrum limits. Using eqs.~(\ref{densitypara}) 
and (\ref{variancefinal}) we can translate the UCMH abundance constraints
into constraints on the amplitude of the spectrum of the curvature perturbation.
For each UCMH mass considered we take the pivot point,
$k_0$, to correspond to the length scale of the perturbation (see
eq.~(\ref{smoothingscale})) which eventually forms the UCMH,
$k_0=1/R$.

\begin{figure}
\includegraphics[width=9cm]{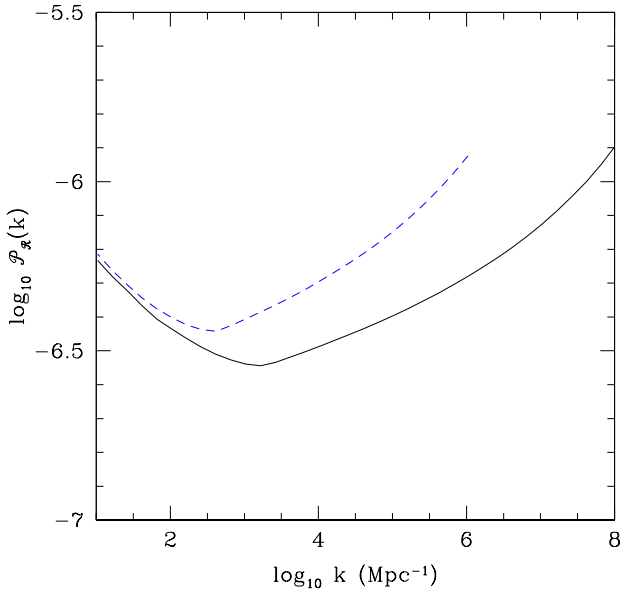}
\caption{Limits on the power spectrum of the primordial curvature
  perturbation as a function of comoving wavenumber (in units of
  ${\rm{Mpc}}^{-1}$).  The solid line shows the potential 
  lower bound on the power spectrum resulting from the detection 
  of gamma-rays from an UCMH by {\em{Fermi}} at threshold sensitivity.  The dotted  line 
  shows the upper limit on the power spectrum obtained if gamma-rays 
  from UCMHs are not detected by {\em{Fermi}}, assuming DM is in the form of WIMPs
  and UCMHs are not disrupted during structure formation.}
\label{results2}
\end{figure}

Fig.~\ref{results2} shows the constraints on the power spectrum of the primordial curvature perturbation.
The potential lower limit on the power spectrum which would arise from the detection of gamma-rays by {\em{Fermi}} from a single UCMH is of the order $\mathcal{P_R} \gtrsim 10^{-6.6}-10^{-5.9}$ on scales $k \sim 10^{1}-10^{8} \, {\rm Mpc}^{-1}$. If gamma-ray emission from UCMHs is not observed, an upper limit can be placed on the power spectrum of the primordial curvature
perturbation, of the order $\mathcal{P_R}  \lesssim 10^{-6.5}-10^{-6}$ on scales
$k \sim 10^{1}-10^{6} \, {\rm Mpc}^{-1}$.
Constraints for larger wavenumbers than those shown in Fig.~\ref{results2} 
result in $f_{\rm UCMH} \gtrsim 1$ and so are not considered.
The lower bound based on a detection at {\em{Fermi}} threshold sensitivity is a conservative
limit (provided that the effects of adiabatic contraction are
insignificant).  The upper limit from non-detection relies on several
assumptions, however, most significantly that the DM is in the form of
WIMPs and that significant disruption to UCMHs does not occur.  If
multiple UCMHs were detected by {\em Fermi} (or ACTs), or the flux was
significantly above the detection threshold, then this would imply a
larger UCMH halo fraction, and hence the lower limits on the power
spectrum of the primordial curvature perturbation would be stronger.

\section{Conclusions}
\label{sectionconclude}

Large amplitude ($10^{-3} - 10^{-1}$) density perturbations can seed
the formation of dense UCMHs~\cite{Ricotti:2009bs}.  If DM is
in the form of WIMPs, WIMP annihilation within UCMHs may produce a
detectable flux of gamma-rays~\cite{Ricotti:2009bs,Scott:2009tu}.  We
have investigated the implications of detection, or non-detection, of gamma-ray emission
from UCMHs by {\em{Fermi}} for the power spectrum of the primordial curvature perturbation. 
We find that detection by {\em{Fermi}} at threshold sensitivity would place a lower limit on the UCMH halo fraction.  The smallest detectable UCMH halo fraction, $f_{\rm UCMH} \gtrsim 10^{-7}$,
is for $M_{\rm UCMH} \sim 10^{3} M_{\odot}$.

If {\em{Fermi}} does not detect gamma-ray emission from UCMHs then, assuming DM is in the form of WIMPs, this would place an upper limit on the UCMH halo fraction. The limit is tightest, $f_{\rm UCMH} \lesssim  10^{-5}$,
for $M_{\rm UCMH} \sim 10^{5} M_{\odot}$.  The resulting potential upper limit on the power spectrum of the primordial curvature perturbation, assuming UCMHs are not disrupted during structure formation, would be $\mathcal{P_R} \lesssim  10^{-6.5}-10^{-6}$ on scales $k \sim 10^{1}-10^{6} \, {\rm Mpc}^{-1}$.  This upper bound is significantly stronger than those from primordial black hole formation, $\mathcal{P_R} \lesssim 10^{-1}-10^{-2}$~\cite{Josan:2009qn}, and would hence provide a tighter constraint on models of inflation (c.f. Ref.~\cite{Peiris:2008be,Josan:2010cj}).
It does, however, rely on the assumptions that DM is in the form of WIMPs and UCMHs are not disrupted during the formation of the Milky Way halo.

\begin{acknowledgments}
We are grateful to Patt Scott for
useful discussions. AJ is supported by the University of Nottingham, and 
AMG by STFC.
\end{acknowledgments}

\bibliography{ucmhrev}

\bibliographystyle{aa}

\end{document}